\documentclass[aps,prl,twocolumn,showpacs]{revtex4}

\usepackage{bm}%
\usepackage[colorlinks=true,linkcolor=red]{hyperref}
\usepackage{amsmath}
\usepackage{amsfonts}
\usepackage{amssymb}
\usepackage{verbatim}
\usepackage{epsfig}
\usepackage{graphicx}
\usepackage[sort&compress]{natbib}

\def\Tr{\mathrm{Tr}}
\begin{document}

\title{\bf\noindent Distributions of Conductance and Shot Noise and
Associated Phase Transitions}
\author{Pierpaolo Vivo$^1$, Satya N. Majumdar$^2$ and Oriol Bohigas$^2$}
\affiliation{
$^1$ Department of Mathematical Sciences,Brunel
University, Uxbridge, Middlesex, UB8 3PH (United Kingdom) \\
$^2$ Laboratoire de Physique Th\'{e}orique et Mod\`{e}les
Statistiques (UMR 8626 du CNRS), Universit\'{e} Paris-Sud,
B\^{a}timent 100, 91405 Orsay Cedex (France) }

\date{\today}

\begin{abstract}
For a chaotic cavity with two indentical leads each supporting $N$ channels,
we
compute analytically, for large $N$, the full distribution of the conductance and
the shot noise power and show that in both cases there is a
central Gaussian region flanked on both sides by non-Gaussian
tails. The distribution is weakly singular at the junction of
Gaussian and non-Gaussian regimes,  a direct consequence of two
phase transitions in an associated Coulomb gas problem.
\end{abstract}

\pacs{73.23.-b, 02.10.Yn, 24.60.-k, 21.10.Ft}

\maketitle

Ballistic transport of electrons through a mesoscopic cavity such as a
quantum dot has been studied intensively both theoretically and experimentally
in recent times~\cite{beenakker}. In the simplest setting one considers
a single cavity of submicron dimension (e.g., a quantum dot) with two identical leads
(each supporting $N$ channels)
connecting it to two separate electron reservoirs. An electron, injected through
one lead, gets scattered in the cavity and leaves by either of the two leads.
The transport of electrons through such an open quantum system is encoded in the
$2N\times 2N$ unitary scattering matrix $ S=
  \begin{pmatrix}
    r & t^\prime \\
    t & r^\prime
  \end{pmatrix}$ connecting the
incoming and outgoing electron wavefunctions, where $r,t$ are
$N\times N$ reflection and transmission matrices from the left and
$r^\prime,t^\prime$ from the right. Several experimentally
measurable transport observables such as the conductance (i.e.,
the time averaged current) and the shot noise power (that
describes the current fluctuations associated with the granularity
of electronic charge $e$) can be expressed in terms of $N$
transmission eigenvalues $T_i$'s of the $N \times N$ matrix $T=t
t^{\dagger}$. For example, the dimensionless conductance $G= \Tr(t
t^\dagger)=\sum_{i=1}^N T_i$ \cite{landauer} and the shot noise
$P=\Tr[t t^\dagger(1-t t^\dagger)]=\sum_{i=1}^N T_i(1-T_i)$
\cite{ya}. The eigenvalue $0\le T_i\le 1$ has a simple physical
interpretation as the probability that an electron gets
transmitted through the $i$-th channel.

Over the past two decades, the random matrix theory (RMT) has been
successfully used~\cite{beenakker,baranger} to model the transport
through such a cavity. Within this RMT approach, one assigns a
uniform probability density to all $S$ matrices belonging to the
unitary group: this amounts to drawing $S$ at random from one of
Dyson's Circular Ensembles \cite{MehtaCond}. This, in turn,
induces a probability measure over the transmission eigenvalues
$T_i$'s whose joint probability density (jpd) is known
\cite{beenakker,baranger,forrcond}
\begin{equation}
P\left(\{T_i\}\right)=A_N\, \prod_{j<k}|T_j-T_k|^\beta\, \prod_{i=1}^N T_i^{\beta/2-1}
\label{jpd1}
\end{equation}
where $A_N^{-1}= \int_0^1\ldots \int_0^1 P\left(\{T_i\}\right)
\prod_i dT_i$ is a normalization constant and the Dyson index
$\beta$ characterizes different symmetry classes ($\beta=1,2$
according to the presence or absence of time-reversal symmetry and
$\beta=4$ in case of spin-flip symmetry). Given this jpd in
\eqref{jpd1} one is then naturally interested in computing the
statistics of transport observables such as the conductance $G$
and the shot noise $P$.

The average and variance of $G$ and $P$ are
known~\cite{beenakker,ya} and in particular for large $N\gg 1$,
$\langle G\rangle=N/2$, $\langle P\rangle=N/8$,
$\mathrm{var}(G)=1/8\beta$, $\mathrm{var}(P)=1/64\beta$. The fact
that the variance of $G$ becomes independent of $N$ for large $N$
has been dubbed the `universal conductance
fluctuations'~\cite{beenakker}. In contrast, much less is known
for their full distributions. For the conductance, the full
distribution is known explicitly only for $N=1$ and
$N=2$~\cite{baranger} and very recently, a solution in terms of
Painlev\'{e} trascendents for the special case $\beta=2$ has been
announced \cite{kanzosip}. Also, the behavior of these
distributions very close to the endpoints (e.g., near $G=0$ and
$G=N$) have been computed recently~\cite{sommers}.

The purpose of this Letter is to present exact results for the full distributions
$\mathcal{P}(G,N)$ and ${\mathcal P}(P,N)$ for all $\beta$ in the large $N$ limit. For the
conductance $G$, we show that for large $N$, $
\mathcal{P}(G,N)\approx \exp\left(-\frac{\beta}{2}N^2
\Psi_G\left(\frac{G}{N}\right)\right) $
where $\approx$ stands for the precise
asymptotic law:
\begin{equation}\label{PreciseAsymptoticLaw}
  \lim_{N\to\infty}\left[-\frac{2\ln\mathcal{P}(Nx,N)}{\beta
  N^2}\right]=\Psi_G(x)
\end{equation}
and the {\em rate function} $\Psi_G(x)$ is explicitly given in
\eqref{PsiG}. Thus the distribution $\mathcal{P}(G,N)$ has a pure
Gaussian form around the mean $\langle G\rangle =N/2$ over the
region $N/4 \le G \le 3N/4$ and outside this central zone, it has
non-Gaussian large deviation tails. The distribution has an
extraordinarily weak singularity at the points $G=N/4$ and
$G=3N/4$ (only the $3$-rd derivative is discontinuous!). One of
our central results is to show that these two weak singularities
in the conductance distribution are the direct consequence of two
phase transitions in an associated Coulomb gas problem. The
distribution of the shot noise $P$ in the allowed range $0\le P\le
N/4$ displays a similar large $N$ behavior: $
\mathcal{P}(P,N)\approx \exp\left(-\frac{\beta}{2}N^2
\Psi_P\left(\frac{P}{N}\right)\right) $ where $\Psi_P(x)$ is given
in \eqref{rateFano}. Once again the central Gaussian regime around
the mean $\langle P\rangle =N/8$ is flanked by two non-Gaussian
tails with a weak third order singularities at the transition
points $P= N/16$ and $P=3N/16$. In addition to these
distributions, we have also computed exactly the large $N$
statistics of the integer moments $\mathcal{T}_n=\sum_{i=1}^N
T_i^n$. Even though our analytical results are valid for large
$N$, we find, rather remarkably, that the numerical results for
relatively small values of $N$ (such as $N=4$) agree fairly well
with our asymptotic results.

We start by demonstrating how the computation of $\mathcal{P}(G,N)$ can be
mapped to the calculation of the partition function of a Coulomb gas problem.
By definition, $\mathcal{P}(G,N)=\langle \delta\left(G-\sum_i T_i\right)\rangle$
where $\langle \rangle$ denotes an average over the jpd in \eqref{jpd1}.
It is then natural to consider the Laplace transform $\langle e^{-\beta p N G/2}\rangle=
\int {\mathcal P}(G,N) e^{-\beta\, p\, N\, G/2} dG$, where the Laplace variable
$p$ is scaled by the factor $\beta N/2$ for later convenience. In terms of the eigenvalues $T_i$,
this Laplace transform can then be expressed as a ratio of two partition functions
\begin{equation}
\langle e^{-\frac{\beta p N}{2} G}\rangle= \frac{\int e^{-\frac{\beta p N}{2} \sum_i
T_i}\,P\left(\{T_i\}\right)\,
\prod_i {dT_i}}{\int P\left(\{T_i\}\right)\,
\prod_i {dT_i}}= \frac{Z_p(N)}{Z_0(N)}.
\label{ratio1}
\end{equation}
Using \eqref{jpd1}
\begin{equation}
Z_p(N)=\int_{0}^{1}\ldots \int_0^{1}\, \exp\left[-\beta
E\left(\{T_i\}\right)\right]\, \prod_i dT_i
\label{pf1}
\end{equation}
where $E\left(\{T_i\}\right)=\sum_i\left(\frac{pN}{2} T_i -
\frac{(\beta-2)}{2\beta}\ln(T_i)\right)- \sum_{j< k} \ln
|T_j-T_k|$. Thus $T_i$'s can be interpreted as the positions of
charged particles confined in a $1$-d box $[0,1]$, repelling each
other via the $2$-d Coulomb potential (logarithmic) and each
subject to an external potential with a linear (with amplitude
$pN/2$) and a logarithmic part, the latter being subdominant in
the large $N\gg 1$ limit. Then $E$ is the
energy of a configuration and $Z_N(p)$ is the partition function
at an inverse temperature $\beta$.

Our next step is to evaluate the partition function $Z_p(N)$ in the large $N$ limit.
For that one carries out the multiple integral in \eqref{pf1} in two steps.
The first step is a coarse-graining procedure where one sums over (partial tracing) all
microscopic
configurations of $T_i$'s compatible with a fixed charge density function $\varrho_p(T)=N^{-1}\sum_i
\delta(T-T_i)$ and the second step consists in performing a functional integral
over all possible positive charge densities $\varrho_p(T)$ that are normalized to unity.
Finally the functional integral is carried out in the large $N$ limit by the saddle point
method. This procedure has recently been used in the context of the largest eigenvalue distribution of
Gaussian~\cite{DM} and Wishart random matrices~\cite{VMB} and also in other related problems
of counting of stationary points in random Gaussian landscapes~\cite{BrayDean}. Following
the general procedure in ~\cite{DM},
the resulting functional integral, to leading order in large $N$, becomes
\begin{equation}
Z_p(N) \propto \int\mathcal{D}[\varrho_p]e^{-\frac{\beta}{2}N^2 S[\varrho_p]}
\label{pf2}
\end{equation}
where the action is given by
\begin{align}\label{Action Sp}
 \nonumber S[\varrho_p] &=p\int_0^1 \varrho_p(T)T dT+B\left[\int_0^1 \varrho_p(T)dT-1\right]\\
 &-\int_0^1\int_0^1 dTdT^\prime
  \varrho_p(T)\varrho_p(T^\prime)\ln|T-T^\prime|.
\end{align}
Here $B$ is a Lagrange multiplier enforcing the normalization of
$\varrho_p(T)$. In the large $N$ limit,
one then evaluates the functional integral in \eqref{pf2} by the saddle point method, i.e.,
one finds the solution $\varrho_p^\star(T)$ (the equilibrium charge density that minimizes
the action or the free energy) from the stationarity condition $\delta
S[\varrho_p]/\delta\varrho_p=0$ which leads to an
integral equation
$p T + B = 2\int_0^1 \varrho_p^\star (T^\prime)\, \ln|T-T^\prime|\, dT'$. Differentiating
once with respect to $T$ leads to a singular integral equation
\begin{equation}
\frac{p}{2}={\rm Pr}\int_0^1\frac{\varrho_p^\star(T^\prime)}{T-T^\prime}dT^\prime
\label{sing1}
\end{equation}
where ${\rm Pr}$ denotes the principal part. Assuming one can solve \eqref{sing1}
for $\varrho_p^\star$, one next evaluates the action
$S[\varrho_p]$
in \eqref{Action Sp} at the stationary solution $\varrho_p^\star$ and then takes
the ratio in \eqref{ratio1} to get
\begin{equation}
\label{AsymptoticDecay}
\langle e^{-\beta p N G/2}\rangle \approx e^{-\frac{\beta}{2}N^2
[\overbrace{S[\varrho_p^\star]-S[\varrho_0^\star]}^{J_G(p)}]}.
\end{equation}
Inverting the Laplace transform gives the main asymptotic result $
\mathcal{P}(G,N)\approx \exp\left(-\frac{\beta}{2}N^2
\Psi_G\left(\frac{G}{N}\right)\right) $ where the rate function
is a Legendre transform,
$\Psi_G(x)=\max_p[-xp+J_G(p)]$ with $J_G(p)$ given by the {\em free energy difference}
as in \eqref{AsymptoticDecay}.

\begin{figure}[htb]
\centerline{\epsfig{ figure=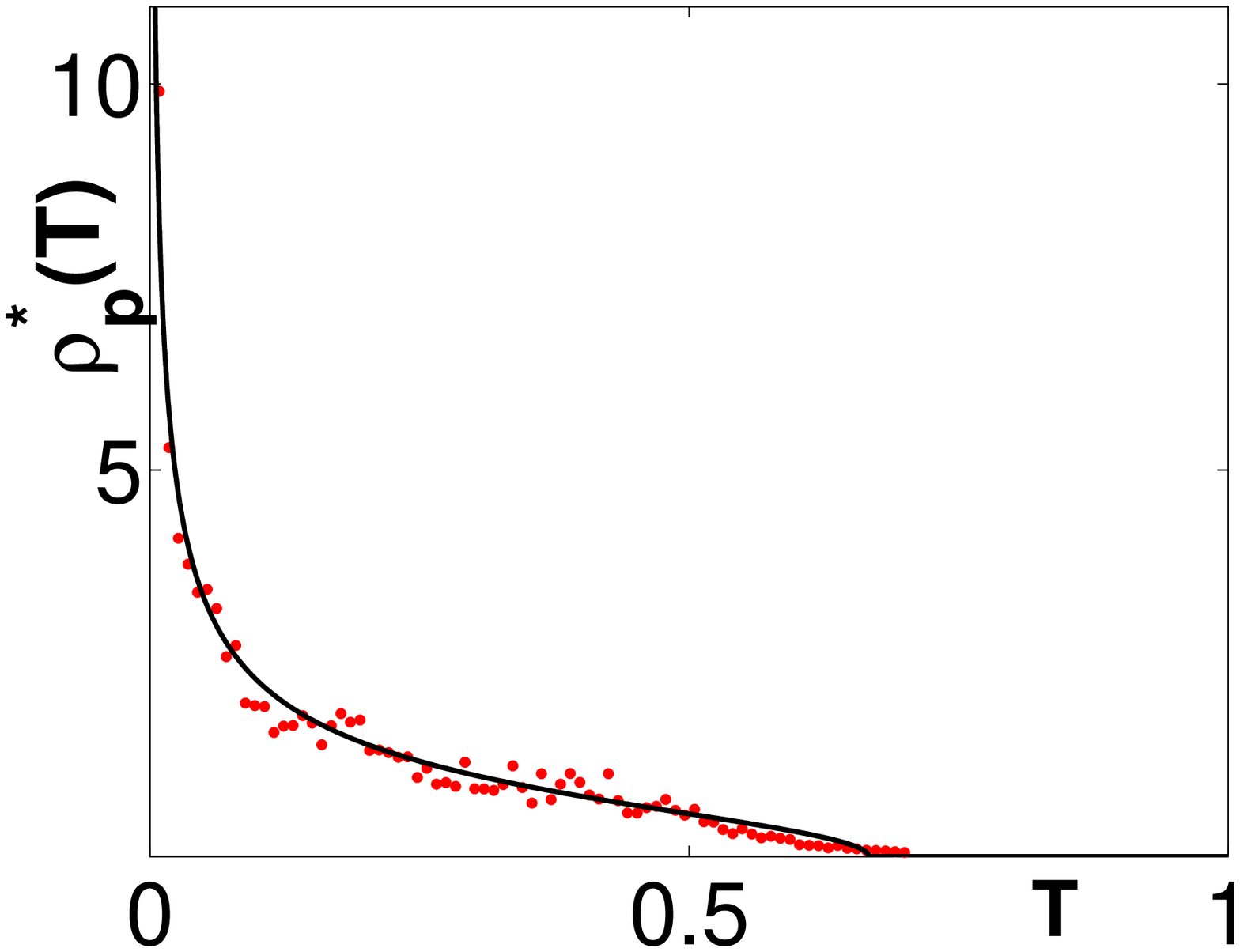 ,width=7pc} \epsfig{
figure=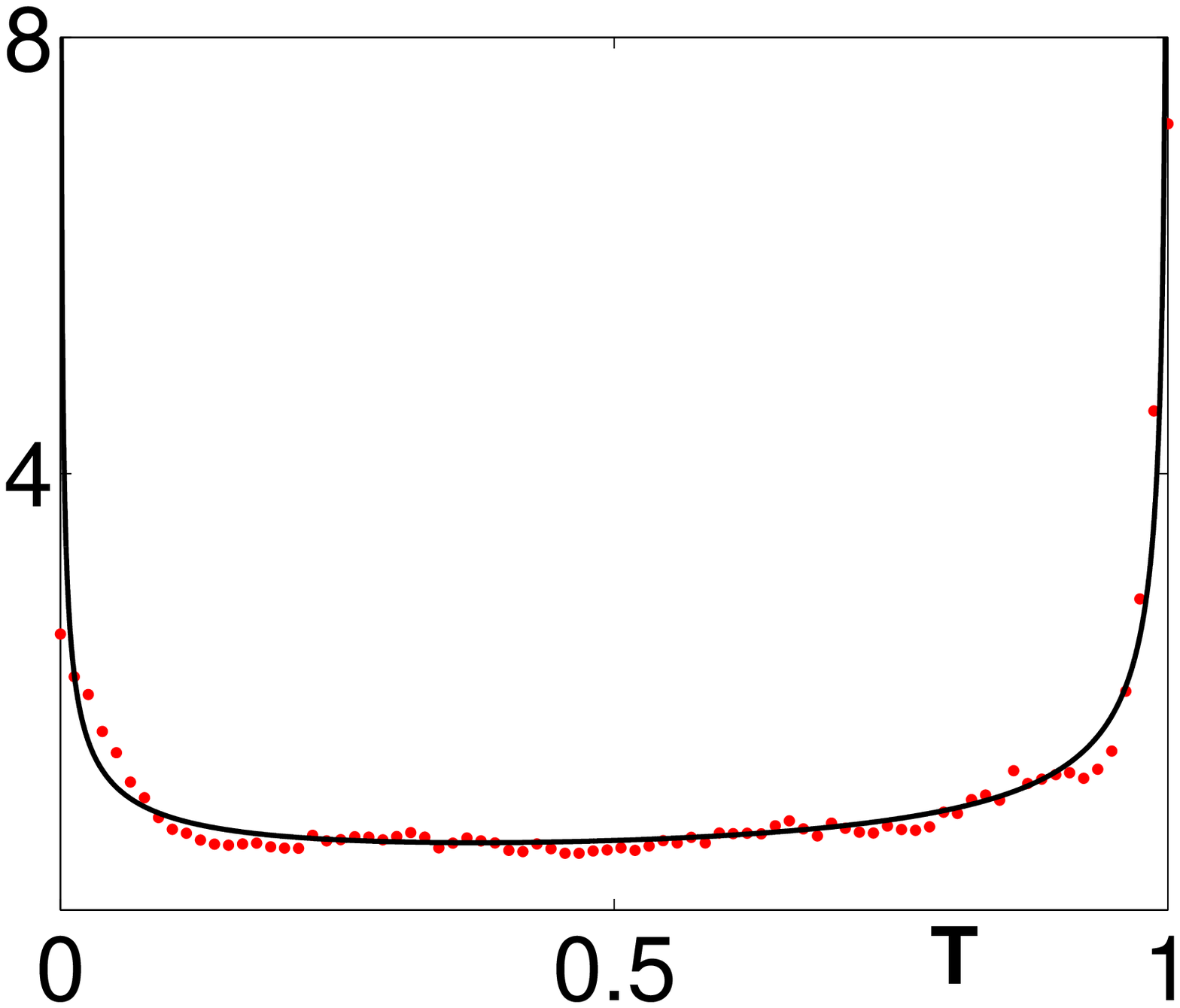 ,width=6.5pc}\epsfig{
figure=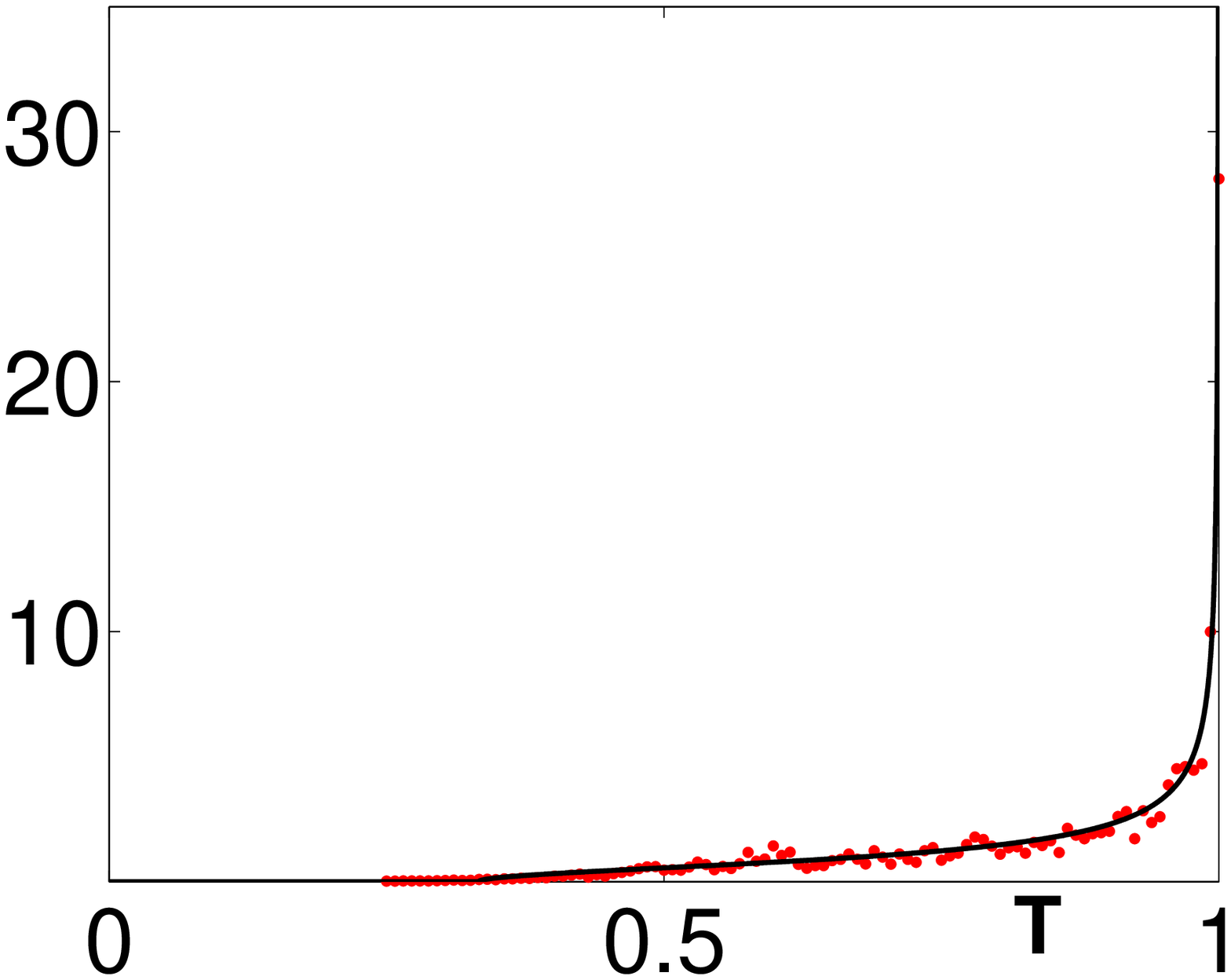,width=6.9pc}} \caption{Charge density
$\varrho_p^\star(T)$ for conductance (numerical, in red dots;
theory, in solid black line). From left to right $p=6,-1,-6$. The
numerical density~\cite{simulation1} is for $N=5$ and $\beta=2$.
\label{DensityCond}}
\end{figure}

It then rests to solve \eqref{sing1} to find the equilibrium
charge density $\varrho_p^\star(T)$. Fortunately, singular
integral equations such as (\ref{sing1}) can be solved in closed
form using Tricomi's theorem~\cite{Tricomi1}. Skipping
details~\cite{details}, we find that the solution
$\varrho_p^\star(T)$ has three different forms (see Fig. 1)
depending on $p$, the parameter that controls the linear external
potential. We get
\begin{displaymath}\label{DensitySummary}
\displaystyle{\varrho_p^\star(T)} =
 \begin{cases}
  \displaystyle\frac{p\sqrt{\frac{4}{p}-T}}{2\pi\sqrt{T}};
  \,\, 0\leq T\leq \frac{4}{p}; \quad p\geq 4\\
  \vspace{2pt}\\
  \displaystyle\frac{p\left[\frac{4+p}{2p}-T\right]}{2\pi\sqrt{T(1-T)}};
  \,\, 0\leq T\leq 1; \quad -4\leq p\leq 4\\
  \vspace{2pt}\\
\displaystyle\frac{|p|\sqrt{T-(1-\frac{4}{|p|})}}{2\pi\sqrt{1-T}};
  \,\, 1-\frac{4}{|p|}\leq T\leq 1; \quad p\leq -4
  \end{cases}
\end{displaymath}
Thus the equilibrium charge density undergoes two {\em phase transitions}
as one tunes $p$ respectively at $p=4$ and $p=-4$. For $p\geq 4$, the
linear potential is strong enough to confine the charges near $T=0$
leading to an inverse square root divergence of the density at $T=0$
and the density becomes zero beyond $4/p$. As $p$ approaches the critical value $4$
from above, the upper edge of the density support approaches the maximum value $1$ and
for $-4\le p \le 4$ the charges are spread over the full box $[0,1]$ with the density
diverging at the two end points. Finally for $p\leq -4$, charges accumulate on the
right wall $T=1$ of the box with the density vanishing for $T\leq 1-4/|p|$.
Note that the equilibrium density (in the large $N$ limit) is independent
of $\beta$.
The analytical solutions in the $3$ regimes above are depicted in Fig. \ref{DensityCond}
together with MC results~\cite{simulation1}.

\begin{figure}[htb]
\centerline{\epsfig{ figure=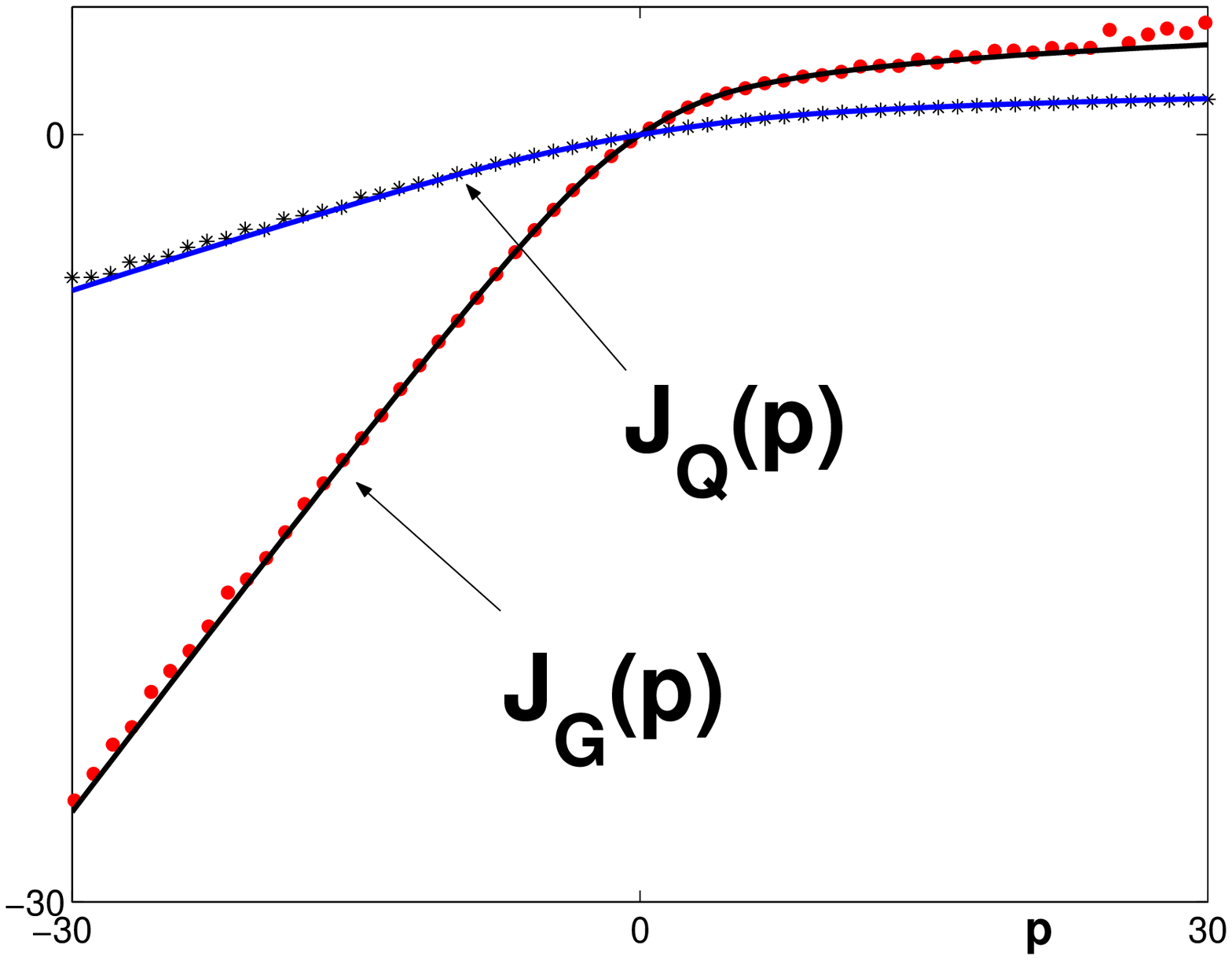 ,width=10pc} \epsfig{
figure=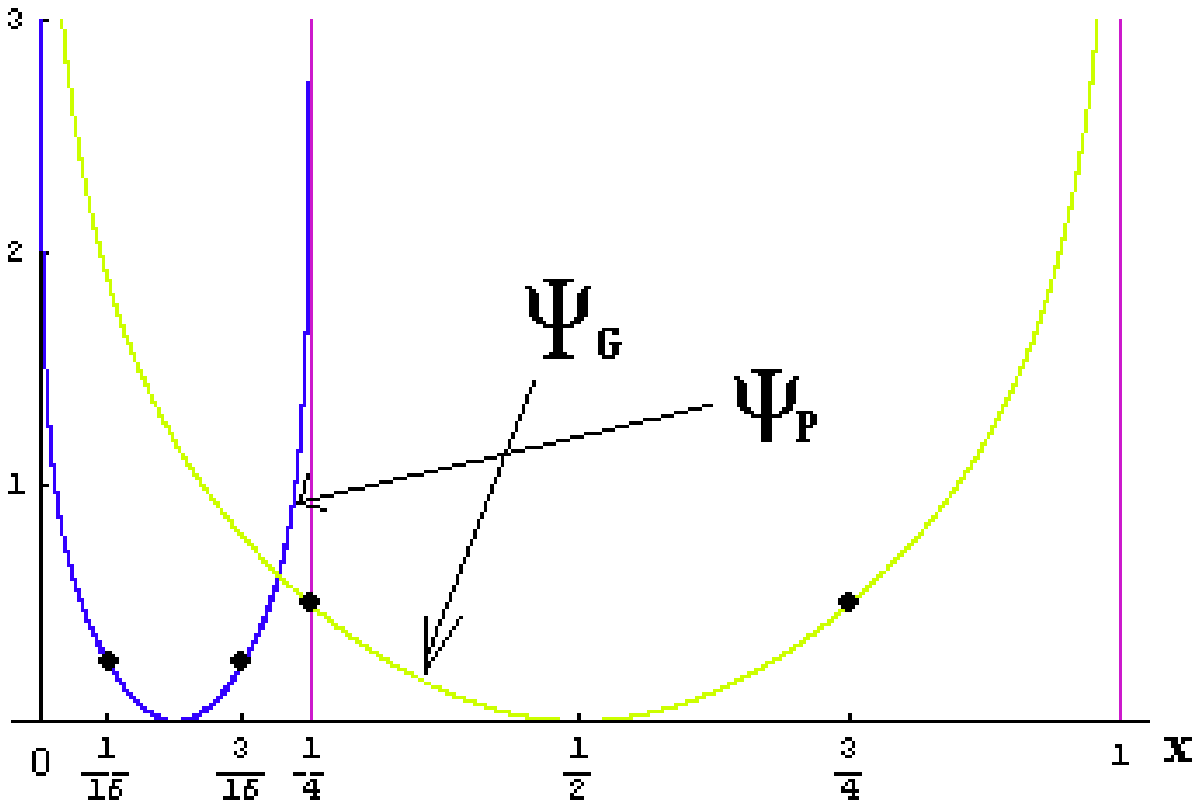 ,width=13pc}} \caption{\textbf{Left:} Free energy
difference $J_G(p)$ (red) and $J_Q(p)$ (blue) from Montecarlo (MC)
simulations and theory (solid line). We have used $-30\le p\le
30$, $\beta=2$ and $N=5$ (for conductance) and $N=4$ (for shot
noise) respectively. \textbf{Right:} Rate functions $\Psi_G(x)$
(green) and $\Psi_P(x)$ (blue) (see \eqref{PsiG} and
\eqref{rateFano}). The black dots indicate the two critical points
on each curve.\label{JGP}}
\end{figure}

Inserting  $\varrho_p^\star(T)$ in the action \eqref{Action Sp} one obtains the
free energy difference in \eqref{AsymptoticDecay}
\begin{equation}\label{JGdiP}
  J_G(p)=
  \begin{cases}
3/2+p + \ln(-p/4) &
  p\leq -4 \\
-\frac{p^2}{32}+\frac{p}{2} &
 -4\leq p\leq 4\\
3/2+\ln(p/4) &
   p\geq 4
  \end{cases}
\end{equation}
This analytical prediction is compared in Fig. \ref{JGP} with MC
simulations performed using similar tricks as in the case of
$\varrho_p^\star(T)$. Finally, the rate function
$\Psi_G(x)=\max_p[-xp+J_G(p)]$ in \eqref{PreciseAsymptoticLaw} is
then given by (see Fig. \ref{JGP}):
\begin{equation}\label{PsiG}
   \Psi_G(x)=
  \begin{cases}
\frac{1}{2}-\ln(4x) & 0\leq x\leq \frac{1}{4}\\
    8\left(x-\frac{1}{2}\right)^2 & \frac{1}{4}\leq x\leq \frac{3}{4} \\
    \frac{1}{2}-\ln[4(1-x)] & \frac{3}{4}\leq x\leq 1
  \end{cases}
\end{equation}
Using the quadratic form in $1/4\le x=G/N \le 3/4$ above in
$\mathcal{P}(G,N)\approx \exp\left(-\frac{\beta}{2}N^2
\Psi_G\left(\frac{G}{N}\right)\right)=
e^{-4\,\beta\,\left(G-\frac{N}{2}\right)^2} $ gives the Gaussian
form in this central region from which one easily reads off the
well known values for mean and variance $\langle G\rangle=N/2$ and
$\mathrm{var}(G)=1/8\beta$. On the other hand, near the two
endpoints $x=0$ and $x=1$, using $\Psi_G(x) \sim -\ln(x)$ and
$\sim -\ln(1-x)$ as in \eqref{PsiG}, we get the power law
dependence, $\mathcal{P}(G,N)\sim G^{\beta N^2/2}$ (as $G\to 0$)
and $\mathcal{P}(G,N)\sim (N-G)^{\beta N^2/2}$ (as $G\to N$) which
are in agreement, to leading order in large $N$, with the exact
far tails obtained in \cite{sommers}. The central Gaussian regime
matches smoothly with the two side regimes (only the third
derivative is discontinuous). Such weak nonanaliticities of third
order were also recently found in the entropy distribution in a
quantum entanglement problem~\cite{facchi}. An expression similar
to \eqref{PsiG} was recently found in \cite{kanzosip} by a
different method. However, the intermediate regime with an
exponential decay claimed in \cite{kanzosip} does not appear in
our solution.

\begin{figure}[htb]
\centerline{\epsfig{ figure=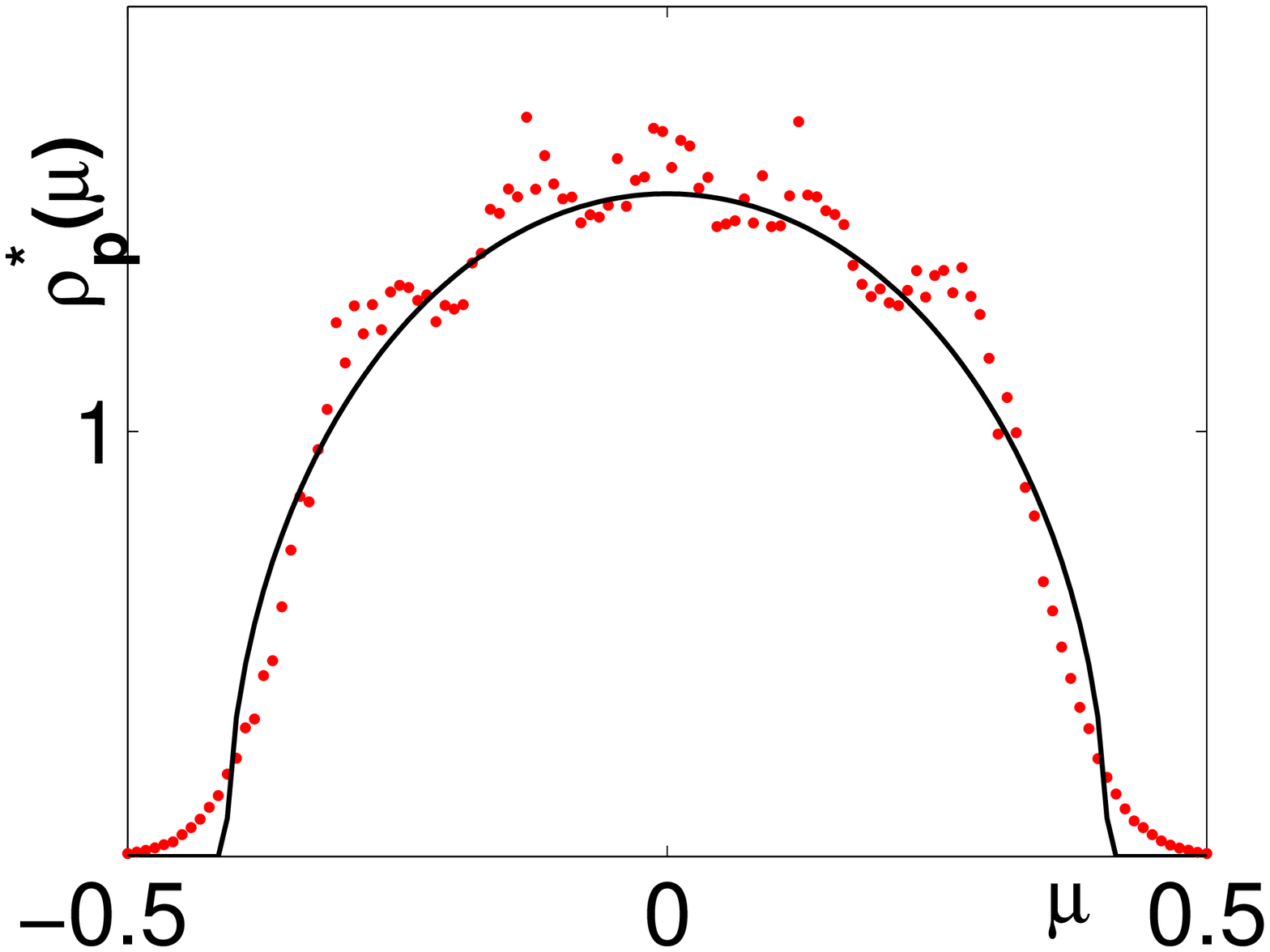 ,width=7.6pc} \epsfig{
figure=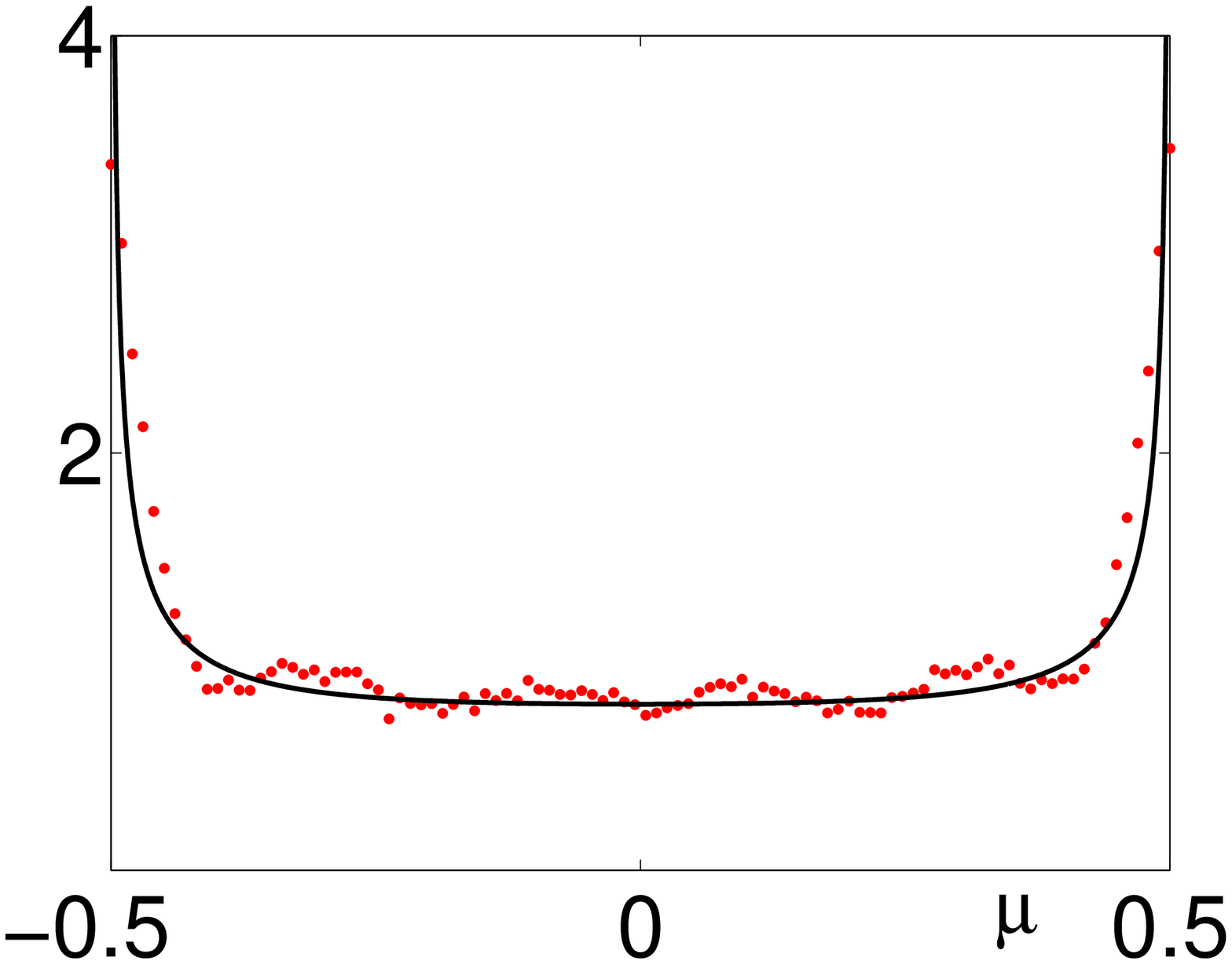 ,width=7.2pc}\epsfig{ figure=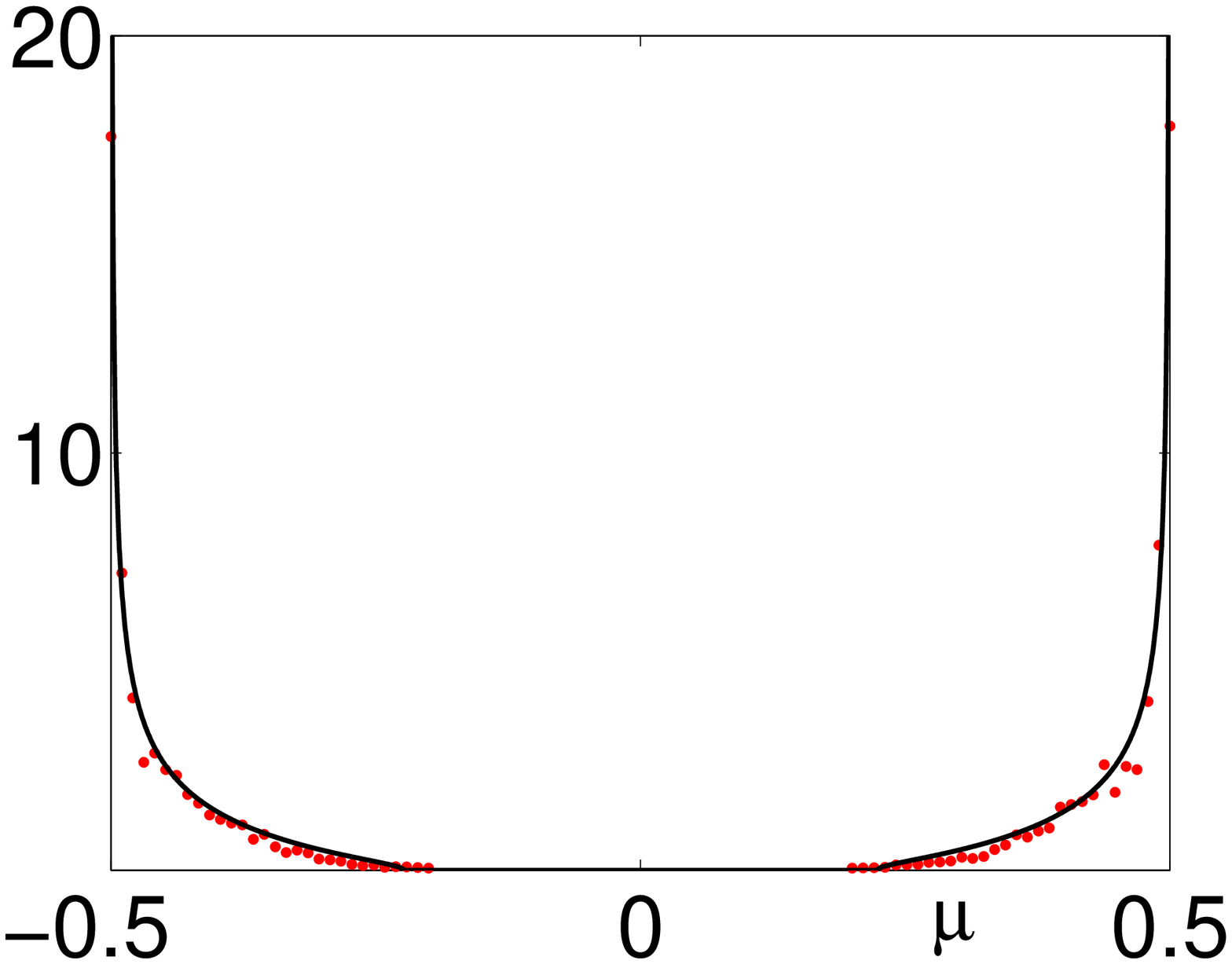
,width=7.2pc}} \caption{Charge density $\varrho_p^\star(\mu)$ for
the shifted shot noise $Q$. From left to right $p=12,2,-10$.
\label{DensityFano}}
\end{figure}

The same formalism can be easily extended to find the distribution
$\mathcal{P}(P,N)$ of the shot noise $P=\sum_{i=1}^N T_i(1-T_i)$.
It turns out to be convenient to work with the shifted variable
$Q= N/4-P= \sum_{i=1}^N (1/2-T_i)^2$. Considering the Laplace
transform of the distribution of $Q$, we then get a new Coulomb
gas where the charges, i.e., the shifted eigenvalues
$\mu_i=1/2-T_i$'s are confined in the box $[-1/2,1/2]$, repel each
other logarithmically and each experiences an external harmonic
(with amplitude proportional to the Laplace variable $p$)
+logarithmic potential. The equilibrium charge density
$\varrho_p^\star(\mu)$ of this Coulomb gas can be computed again
by solving the integral equation $ p\mu={\rm
Pr}\int_{-1/2}^{1/2}\frac{\varrho_p^\star(\mu^\prime)}{\mu-\mu^\prime}d\mu^\prime
$. Skipping details~\cite{details}, we again find two phase
transitions upon tuning the value of $p$. For $p>8$, the density
has a semi-circular form over $\mu\in [-\sqrt{2/p},\sqrt{2/p}]$.
For $-8<p<8$, the density has square root divergences at the box
ends $\mu=\pm 1/2$. For $p<-8$, the density breaks up into two
disjoint sectors around the two endpoints and vanishes in the
middle of the box (see Fig. \ref{DensityFano}). In these three
regions $p>8$, $-8<p<8$ and $p<-8$ the exact form of the density
is given by
\begin{displaymath}\label{DensitySummaryFano}
\displaystyle{\varrho_p^\star(\mu)}= \begin{cases}
  \displaystyle\frac{p}{\pi}\sqrt{\frac{2}{p}-\mu^2};
  \quad -\sqrt{\frac{2}{p}}\leq \mu\leq \sqrt{\frac{2}{p}} \\
 \vspace{2pt}\\
\displaystyle\frac{p}{\pi\sqrt{1/4-\mu^2}}\left[\frac{8+p}{8p}-\mu^2\right];
\quad   -\frac{1}{2}\leq \mu\leq \frac{1}{2} \\
\vspace{2pt}\\
\displaystyle\frac{|p\mu|\sqrt{\mu^2-\zeta_p^2}}{\pi\sqrt{1/4-\mu^2}};\quad
 |\zeta_p| \leq |\mu|\leq \frac{1}{2}
  \end{cases}
\end{displaymath}
where $\zeta_p=\sqrt{1/4-2/|p|}$ and clearly $|\zeta_p|< 1/2$ for
$p<-8$. In the last case, $\varrho_p^\star(\mu)=0$ for $-\zeta_p<
\mu < \zeta_p$.

Following steps similar to the conductance case, we then compute
the free energy difference associated with the shifted shot noise
$Q=N/4-P$
\begin{equation}\label{JQdiPsummary}
  J_Q(p)=
  \begin{cases}
\frac{3}{4}+\frac{1}{2}\ln \left(\frac{p}{8}\right) &
 p\geq 8\\
\frac{-p^2}{256}+\frac{p}{8} &
   -8\leq p\leq 8\\
  \frac{3-|p|}{4}+\frac{1}{2} \ln \left(\frac{|p|}{8}\right) &
  p\leq -8
  \end{cases}
\end{equation}
from which the rate function for the unshifted shot noise
$P=\sum_i T_i(1-T_i)$ can be computed. Noting that $0\le T_i\le 1$
implies that the scaled shot noise $0\le x=P/N\le 1/4$, we get
(see Fig. \ref{JGP}):
\begin{equation}\label{rateFano}
\Psi_P(x) =
  \begin{cases}
     \frac{1}{4}-\frac{1}{2}\ln(16x)  & 0\leq x\leq \frac{1}{16}\\
    64\left(\frac{1}{8}-x\right)^2  & \frac{1}{16}\leq x\leq \frac{3}{16} \\
    \frac{1}{4}-\frac{1}{2}\ln\left[16\left(\frac{1}{4}-x\right)\right] &  \frac{3}{16}\leq x\leq
\frac{1}{4}.
  \end{cases}
\end{equation}
Once again, the rate function $\Psi_P(x=P/N)$ is weakly
nonanalytic at the critical values $x=P/N=1/16, 3/16$ where a
central Gaussian regime connects the two non-Gaussian regimes on
either sides. From the central regime, one again reads off the
mean and the variance of the shot noise for large $N$: $\langle
P\rangle=N/8$ and $\mathrm{var}(P)=1/64\beta$. At the two
endpoints $P\to 0$ and $P\to N/4$, we get using \eqref{rateFano},
power law behavior for the distribution, $\mathcal{P}(P,N)\sim
P^{\beta N^2/4}$ (as $P\to 0$) and $\mathcal{P}(P,N)\sim
(N/4-P)^{\beta N^2/4}$ (as $P\to N/4$). The latter limit is in
agreement with the far tail result of \cite{sommers}, while the
precise asymptotics near $P\to 0$ was not known before.

The large-$N$ behavior \eqref{PreciseAsymptoticLaw} is generic for
the distribution of any linear statistics on the transmission
eigenvalues (of the form $\sum_i a(T_i)$ where $a(T)$ can be
nonlinear as in the case of shot noise) and the large-$N$
formalism developed here may, in principle, be extended to compute
the distribution of any arbitrary linear statistic $a(T)$. In
particular, as a byproduct one can derive the mean and variance of
any linear statistics by computing the central quadratic behavior
of the associated rate function $\Psi(x)$. In certain cases this
provides new explicit formulae for mean and variance that are not
easily accessible by other general methods such as in
\cite{beenakkerPRL}. As an example we consider briefly the case of
integer moments $\mathcal{T}_n=\sum_{i=1}^N T_i^n$, relevant for
statistics of charge cumulants \cite{sommers,vivovivo,novaes}.
The rate function close to the maximum (Gaussian regime) is given
by \cite{details}: $
  \Psi_{\mathcal{T}_n}(x)=(4 A_n)^{-1}(x-B_n)^2\qquad \text{for }
  x_n^{(-)}<x<x_n^{(+)}
$, where $x_n^{(\pm)}$ are $n$-dependent edge points and the
constants $A_n$ and $B_n$ are given by
$A_n=(2n-1)\Gamma(n+1/2)\Gamma(n-1/2)/16\pi n \Gamma^2(n)$ and
$B_n=4^{-n}\binom{2n}{n}$.

This implies: $ \mathcal{P}(\mathcal{T}_n,N) \approx
\exp\left[-\frac{\beta}{8A_n}
\left(\mathcal{T}_n-B_nN\right)^2\right]$ near the mean. From this
Gaussian form one can read off the mean and the variance. For the
mean we find a $\beta$-independent result
$\langle\mathcal{T}_n\rangle =B_n N =\frac{N}{4^n}\binom{2n}{n}$.
In the special case $\beta=2$ this was recently proved in
\cite{novaes} by other methods. For the variance we find a new
exact result: $ \mathrm{var}(\mathcal{T}_n) = \frac{4 A_n}{\beta}
=
\frac{1}{4\beta\pi}\frac{(2n-1)\Gamma(n+1/2)\Gamma(n-1/2)}{n\Gamma^2(n)}$
which reduces to the conductance result $1/8\beta$ for $n=1$ and
approaches to the universal constant $1/{2\beta \pi}$ as $n\to
\infty$. Our new result for $ \mathrm{var}(\mathcal{T}_n)$ for
$N\gg 1$ and its universal asymptote cannot be obtained easily
from the formula in \cite{novaes}, valid for a finite number of
open channels.

In summary, we have obtained exact analytical distributions for
the conductance and the shot noise for a mesoscopic cavity with
two leads and $N$ channels each in the large $N$ limit. Our
results reveal a rich thermodynamic behavior including two phase
transitions in an associated Coulomb gas problem leading to
extraordinarily weak nonanaliticities in the distributions. We
believe that the Coulomb gas method (well suited for large $N$)
used here is very general and will be useful in other problems as
well.

\textit{Acknowledgments:} PV acknowledges support from Marie Curie
Early Stage Fellowship (NET-ACE project). Helpful discussions with
M. Novaes, D. Savin, G. Akemann and L. Shifrin are also gratefully
acknowledged.


\begin{thebibliography}{99}
\bibitem{simulation1} Noting that the average density is the same as the single
point marginal of a joint density of $N$ variables, it can be expressed as
the ratio (for simplicity we choose $\beta=2$)
$  \varrho_p^\star(x_1)= \Big\langle e^{-pN\sum_{i=1}^N
  x_i}\prod_{j<k}|x_j-x_k|^2\Big\rangle_{N-1}\Big/\Big\langle e^{-pN\sum_{i=1}^N
  x_i}\prod_{j<k}|x_j-x_k|^2\Big\rangle_N
$ where the average is taken over $N-1$ random numbers
${x_2,\ldots,x_N}$ (numerator) and $N$ (denominator) with a flat
measure over $[0,1]$, with $x_1$ spanning the interval $(0,1)$. The
numerator and the denominator are then separately evaluated numerically.
\end{thebibliography}
\end{document}